\documentclass[conference]{IEEEtran}

\usepackage[utf8]{inputenc}
\IEEEoverridecommandlockouts
\usepackage[numbers,sort]{natbib}
\usepackage{amsmath,amssymb,amsfonts}
\usepackage{algorithmic}
\usepackage{graphicx}
\usepackage{textcomp}
\usepackage{xcolor}
\usepackage{gensymb}
\usepackage{array}
\usepackage{url}
\usepackage{threeparttable}
\def\BibTeX{{\rm B\kern-.05em{\sc i\kern-.025em b}\kern-.08em
    T\kern-.1667em\lower.7ex\hbox{E}\kern-.125emX}}

\begin{document}

\title{Are anonymity-seekers just like everybody else? \\ 
    An analysis of contributions to Wikipedia from Tor}

\makeatletter
\newcommand{\linebreakand}{%
  \end{@IEEEauthorhalign}
  \hfill\mbox{}\par
  \mbox{}\hfill\begin{@IEEEauthorhalign}
}
\makeatother

\author{\IEEEauthorblockN{Chau Tran}
\IEEEauthorblockA{\textit{Department of Computer Science \& Engineering} \\
\textit{New York University}\\
New York, USA \\
chau.tran@nyu.edu}
\and
\IEEEauthorblockN{Kaylea Champion}
\IEEEauthorblockA{\textit{Department of Communication} \\
\textit{University of Washington}\\
Seatle, USA \\
kaylea@uw.edu}
\and
\IEEEauthorblockN{Andrea Forte}
\IEEEauthorblockA{\textit{College of Computing \& Informatics} \\
\textit{Drexel University}\\
Philadelphia, USA\\
af468@drexel.edu}

\linebreakand

\IEEEauthorblockN{Benjamin Mako Hill}
\IEEEauthorblockA{\textit{Department of Communication} \\
\textit{University of Washington}\\
Seatle, USA \\
makohill@uw.edu}
\and
\IEEEauthorblockN{Rachel Greenstadt}
\IEEEauthorblockA{\textit{Department of Computer Science \& Engineering} \\
\textit{New York University}\\
New York, USA \\
greenstadt@nyu.edu}
}

\maketitle

\begin{abstract}
User-generated content sites routinely block contributions from users of privacy-enhancing proxies like Tor because of a perception that proxies are a source of vandalism, spam, and abuse. Although these blocks might be effective, collateral damage in the form of unrealized valuable contributions from anonymity seekers is invisible. One of the largest and most important user-generated content sites, Wikipedia, has attempted to block contributions from Tor users since as early as 2005. We demonstrate that these blocks have been imperfect and that thousands of attempts to edit on Wikipedia through Tor have been successful. We draw upon several data sources and analytical techniques to measure and describe the history of Tor editing on Wikipedia over time and to compare contributions from Tor users to those from other groups of Wikipedia users. Our analysis suggests that although Tor users who slip through Wikipedia's ban contribute content that is more likely to be reverted and to revert others, their contributions are otherwise similar in quality to those from other unregistered participants and to the initial contributions of registered users.
\end{abstract}


\section{Introduction}

When a Wikipedia reader using the Tor Browser notices a stylistic error or missing fact and clicks the ``Edit'' button to fix it, they see a message like the one reproduced in \figurename~\ref{fig:wikipedia}. Wikipedia informs would-be Tor contributors that they, like others using open proxy systems to protect their privacy, have been preemptively blocked from contributing. Wikipedia is not alone in the decision to block participation from anonymity-seeking users. Although service providers vary in their approaches, users of privacy-enhancing technologies are unable to participate in a broad range of online experiences \cite{Afroz:2016:NDSS}.


In this work, we attempt to measure the value of contributions made by the privacy-seeking community and compare these contributions to those by other users. We focus on the users of a single service, Wikipedia, and a single privacy-protecting technology, Tor, to understand what is lost when a user-generated content site systematically blocks contributions from users of privacy-enhancing technologies.

\begin{figure}[t]
 \centering

  \includegraphics[width=8cm]{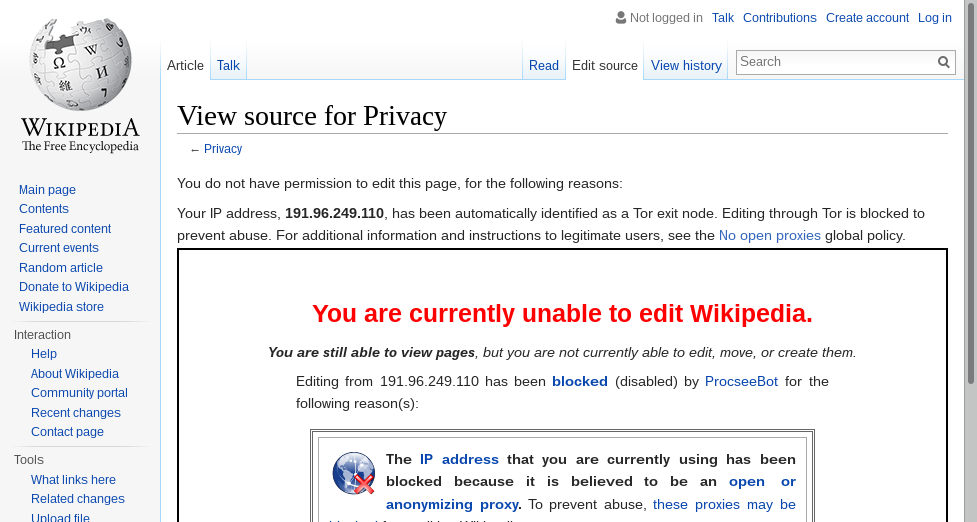}
  \caption{Screenshot of the page a user is shown when they attempt to edit the Wikipedia article on ``Privacy'' while using Tor.}
 \label{fig:wikipedia}
\end{figure}

In particular, we make use of the fact that Wikipedia's mechanism of blocking Tor users has been imperfect to identify and extract 11,363 edits made by Tor users to English Wikipedia between 2007 and 2018. We analyze how some Tor users managed to slip through Wikipedia's block and describe how we constructed our dataset of Tor edits. We use this dataset to compare the edits of people using Tor (\textit{Tor editors}) with three different control sets of time-matched edits by other Wikipedia contributor populations:  (1) non-logged in users editing from non-Tor IP addresses whose edits are credited to their IP address (\textit{IP editors}), (2) people logged into accounts making their first edit (\textit{First-time editors}), and (3) people logged into accounts with more than one edit using the same account (\textit{Registered editors}).

Using a combination of quantitative and qualitative techniques, we find that while Tor editors are more likely to revert someone's work and to be reverted, other indicators of quality suggest that their contributions are similar to those of IP editors and First-time editors. In an exploratory analysis, we fit topic models to Wikipedia articles and find intriguing differences between the kinds of topics that Tor users and other Wikipedia editors contribute to. We conclude with a discussion of how user-generated sites like Wikipedia might accept contributions from the millions of daily users of privacy-enhancing technologies like Tor\footnote{\url{https://metrics.torproject.org/userstats-relay-country.html} (Archived: \url{https://perma.cc/B5W4-UG7C})} in ways that benefit both the websites and society. 

\section{Related Work}

Most people seek out anonymity online at some time or another \cite{donath2002identity}. Their reasons for doing so range from seeking help and support \cite{Andalibi:2016:USM:2858036.2858096}, exploring or disclosing one's identity \citep{Turkle:1995:LSI:526517, doi:10.1111/1540-4560.00247, anonymityeffect}, protecting themselves when contributing to online projects \cite{Forte2017}, seeking information, pursuing hobbies, and engaging in activities that may violate copyright such as file sharing \cite{kang2013people}. 

Anonymity can confer important benefits, not just for the individual seeking anonymity but also for the collective good of online communities \citep{Omernick, kang2013people}. The use of anonymity in collaborative learning has been demonstrated to improve equity, participation rates, and creative thinking \cite{chester}. Research suggests that anonymity can support self-expression and self-discovery among young people \cite{ellison_question_2016}. For instance, researchers found that anonymity helps users discuss topics that are stigmatized \cite{Choudhury, Andalibi:2016:USM:2858036.2858096}. 

Despite the range of legitimate reasons that people adopt anonymity to interact on the Internet and the benefits to collaborative communities, many websites systematically block traffic coming from anonymity-seeking users of systems like Tor.\footnote{\url{https://trac.torproject.org/projects/tor/wiki/org/doc/ListOfServicesBlockingTor} (Archived: \url{https://perma.cc/E49X-MBSE})} According to Khattak et al., at least 1.3 million IP addresses blocked Tor at the TCP/IP level as of 2015, and ``3.67\% of the top 1,000 Alexa sites are blocking people using computers running known Tor exit-node IP addresses''~\cite{Afroz:2016:NDSS}.

Of course, websites do not block anonymity tools like Tor for no reason. Research has shown that online anonymity is sometimes associated with toxic behaviors that are hard to control \cite{Lapidot-Lefler}. Traffic analysis of Tor in 2010 found a substantial portion of network activity is associated with peer-to-peer applications such as BitTorrent \cite{chaabane_digging_2010}. Another report made by Sqreen,\footnote{\url{https://blog.sqreen.io/tor-the-good-the-bad-and-the-ugly/} (Archived: \url{https://perma.cc/38RG-R8JG})} an application protection service, claims that ``a user coming from Tor is between six and eight times more likely to perform an attack'' on their website, such as path scanning and SQL/NoSQL injection. Tor exit node operators often receive complaints of ``copyright infringement, reported hacking attempts, IRC bot network controls, and web page defacements'' \cite{Mccoy:2008:SLD:1428259.1428264}. The most frequent complaints about Tor users' negative behavior are DCMA violations, which made up 99.74\% of the approximately three million email complaints sent to exit operators from Torservers.net from June, 2010 to April, 2016 \cite{203856}.

A third perspective suggests that anonymity seeking behavior is neither ``good'' nor ``bad'' and that anonymous users are best understood as largely similar to other users. Studies of anonymous behaviors on Quora have found that answers from anonymous contributors are no worse than answers given by registered users and the only significant difference is that ``with anonymous answers, social appreciation correlated with the answer’s length''  \cite{DBLP:journals/corr/abs-1811-07223}. Furthermore, Mani et al.'s study of the domains visited by Tor users showed that~80\% of the websites visited by Tor users are in the Alexa top one million, giving further evidence that Tor users are similar to the overall Internet population~\cite{Mani:2018:IMC}.

Although the tradeoffs between anonymity's benefits and threats have been investigated and discussed from many perspectives, the question of what value anonymous contributions might bring to contexts where they are disallowed is difficult to answer. How does one estimate the value of something that is not happening? By examining the relatively small number of Tor edits that slipped through Wikipedia's restriction between 2007--2018, we hope to begin doing just that. In the next sections, we explain the context of our data collection and analysis as well as the methods we used to identify a dataset of Wikipedia edits from Tor.

\section{Empirical Context}

\subsection{Tor}
The Tor network consists of volunteer-run servers that allow users to connect to the Internet without revealing their IP address. Rather than users making a direct connection to a destination website, Tor routes traffic through a series of relays that conceal the origin and route of a user's Internet traffic. Within Tor, each relay only knows the immediate sender and the next receiver of the data but not the complete path that the data packet will take. The destination receives only the final relay in the route (called the ``exit node"), not the Tor user's original IP address. The list of all Tor nodes is published so that Tor clients can pick relays for their circuits. This public list also allows the public to determine whether or not a given IP address is a Tor exit node at a given point in time. Some websites, including Wikipedia, use these lists of exit nodes to restrict traffic from the Tor network. 

\subsection{Wikipedia}
As one of the largest peer production websites, Wikipedia receives vast numbers of contributions every day. While Wikipedia is available in many languages, English Wikipedia is the largest edition with the most articles, active users, and viewers.\footnote{\url{https://en.wikipedia.org/wiki/List_of_Wikipedias} (Archived: \url{https://perma.cc/V2UQ-LBCB})} 
As of February 2019, the English language Wikipedia ``develops at a rate of 1.8 edits per second'' with more than 136,000 registered editors who contribute each month.\footnote{\url{https://en.wikipedia.org/wiki/Wikipedia:Statistics} (Archived: \url{https://perma.cc/4WCW-RNSM})} When these registered editors change something, their username is credited with that edit. Wikipedia also allows people to contribute without asking them to sign up or log in. In these cases, the contributor's IP address is credited with the change.

Wikipedia's low barriers to participation have subjected the website to vandalism and poor-quality editing. In Wikipedia, vandalism refers to the deliberate degradation of an article either by removing part of the existing work or adding damaging content. Erasing the full text of articles and adding profanity or racial slurs are common forms of vandalism. The Wikipedia community invests enormous resources into minimizing and mitigating vandalism. Using a combination of bots and humans, the Wikipedia community has developed banning mechanisms to mitigate repeated attempts from individuals who repeatedly sabotage the community's work. For example, if someone is detected vandalizing an article, their account's privilege to edit on Wikipedia might be halted and the IP address of their device might be banned from editing in the future. Of course, this does not stop more tech-savvy saboteurs from using methods to change their online identities and continuing to cause damage \cite{Geiger:2010:WSO:1718918.1718941}.

Skepticism about anonymity-seeking users has been evident from the early years of Wikipedia. In messages from the the archives of Wikipedia's public mailing lists from 2002 and 2004, Wikipedia's founder Jimmy Wales argued that users without accounts should be treated differently and that anonymous users represented a problem for Wikipedia.\footnote{\url{https://lists.wikimedia.org/pipermail/wikien-l/2002-November/000087.html} (Archived: \url{https://perma.cc/6XQ7-SMP8}) \url{https://lists.wikimedia.org/pipermail/wikien-l/2004-February/010659.html} (Archived: \url{https://perma.cc/56TW-85V3})} In 2005, English Wikipedia blocked anonymous users from creating pages.\footnote{\url{https://lists.wikimedia.org/pipermail/wikien-l/2005-December/033880.html} (Archived: \url{https://perma.cc/DRR8-63PT})} Between 2008 and 2013, there was an extended discussion in the Wikipedia community about how to most effectively block contributions by Tor users.\footnote{\url{https://en.wikipedia.org/wiki/Wikipedia_talk:Blocking_policy/Tor_nodes} (Archived: \url{https://perma.cc/UT2L-VF27})}

Conversations in Wikipedia about allowing anonymity-seeking contributors have rarely discussed the benefits that may flow from allowing them. Recent qualitative research has shown that open-content production sites like Wikipedia value certain forms of anonymous contributions because they can lower barriers to participation but rarely consider other reasons that someone might want to be participate anonymously \cite{mcdonald19chi}. Other work has illuminated the reasons that people want to participate anonymously \cite{Forte2017} and the kinds of good-faith contributions they make \cite{champion_forensic_2019}. This work highlights the differences between service providers' perceptions of what anonymity is good for and what contributors think. As part of this conversation, some Wikipedia users have voiced their concern that the blocking of Tor was not justified and suggested that there had been ``no quantitative information about the frequency and size of [problems created by Tor users].''  Although Wikipedia contributors have occasionally discussed lifting the site's ban on Tor in the mailing lists\footnote{\url{https://lists.wikimedia.org/pipermail/wikien-l/2002-November/000087.html} (Archived: \url{https://perma.cc/6XQ7-SMP8})} and the ``Wikipedia talk: Blocking Policy/Tor nodes'' discussion page,\footnote{\url{https://en.wikipedia.org/wiki/Wikipedia_talk:Blocking_policy/Tor_nodes} (Archived: \url{https://perma.cc/SBZ5-BGMP})} the Tor network remains restricted. 
\section{Tor Edits to Wikipedia}
\subsection{Identifying Tor edits}
\label{sec:identifyingtor}

\begin{figure}[t]
\includegraphics[width=8cm, height=8cm]{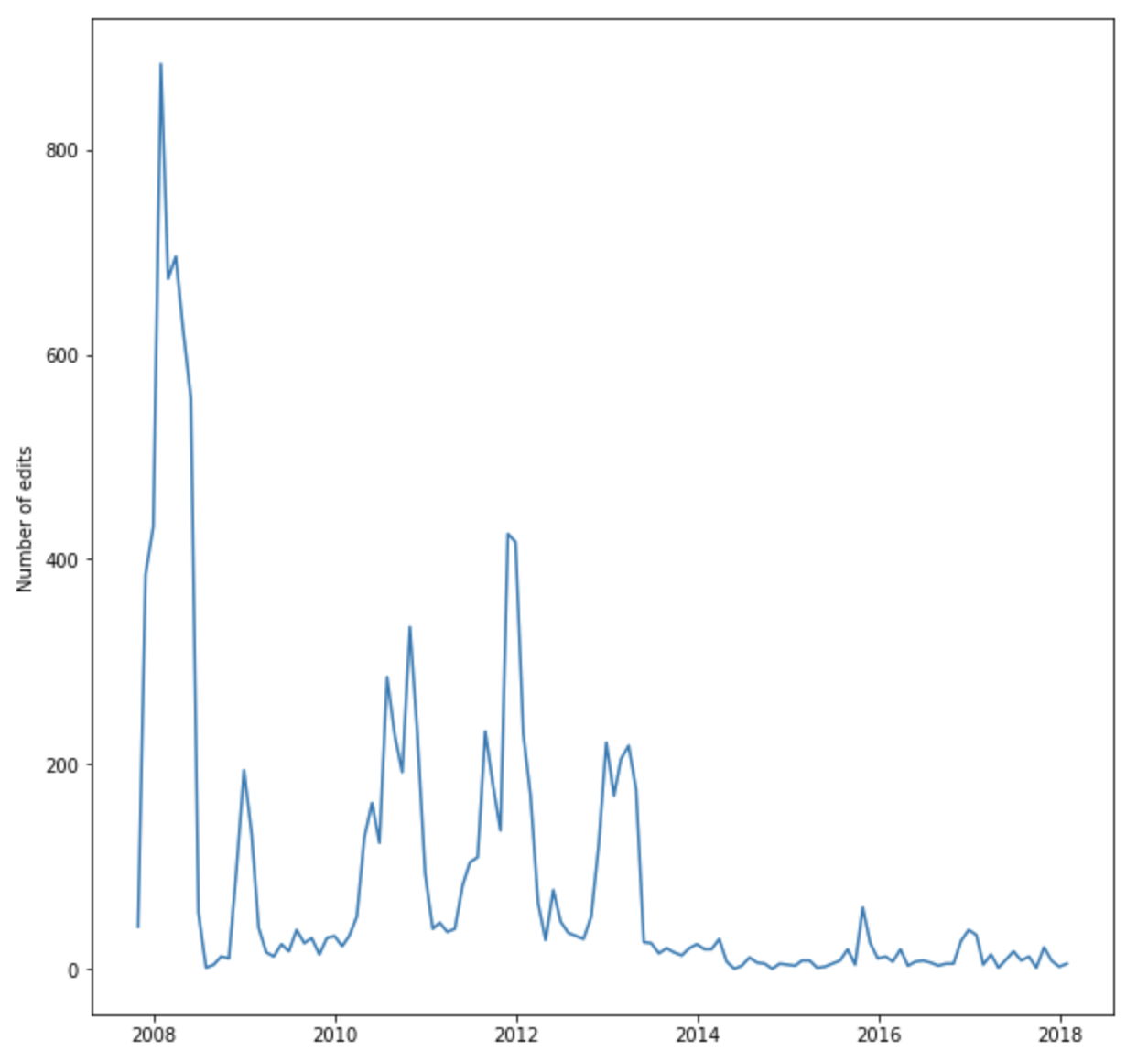}
\caption{Number of edits per month by Tor users to English Wikipedia between 2007 and 2018.}
\label{fig:monthly-edits}
\end{figure}

Edits to Wikipedia made from Tor are attributed to an IP address and appear just like contributions from other unregistered editors. 
To identify edits as coming from Tor, we first used a complete history database dump of the English Wikipedia and obtained metadata of all revisions made on Wikipedia up to March 1, 2018.\footnote{\url{https://dumps.wikimedia.org/} (Archived: \url{https://perma.cc/2G26-G2TJ})} This metadata included revision ID, revision date, editor's username or IP address, article ID or title, and article ``namespace'' (a piece of metadata used to categorize types of pages on Wikipedia). 

The Tor metrics site maintains the list of exit nodes run by volunteers.\footnote{\url{https://metrics.torproject.org/collector.html} (Archived: \url{https://perma.cc/DTC3-TALT})} As the name suggests, the exit list consists of ``known exits and corresponding exit IP addresses available in a specific format.'' Exit list data goes back to February 22, 2010 and is updated and archived every hour. Each archive has details of exit nodes available at the time the list was produced. Most websites that restrict access from Tor, including Wikipedia, have relied on this list. 

When we consulted with the Tor metrics team, we were told that this information is not 100\% complete. Before a node is picked to be an exit node, the Tor network uses dedicated servers to determine whether or not it meets the requirements necessary to function as part of the Tor network. These dedicated servers are called \emph{directory authorities}, and they are in charge of making the available and eligible relays reach a consensus to form a network. Once a consensus is reached, the exit nodes become effective at the time indicated by the directory authorities. This consensus-building process can happen several hours before the exit list is updated. 

As a result, the use of both the consensus and the exit lists is necessary to identify a comprehensive list of exit nodes because sometimes nodes that do not meet the criteria for an exit flag (an identifier flagged by the dedicated server to indicate that a relay is qualified to be an exit node) end up becoming exit nodes anyway due to their exit policy (a set of rules set up by the owner of the relay to dictate how the relay should be operated)~\cite{Afroz:2016:NDSS}. Our dataset of Tor exit nodes reflects a comprehensive set of all exit nodes drawn from both these sources with the specific time periods that the nodes were active.

We crosschecked the IP address and timestamp for every contribution credited to an IP address on Wikipedia to identify any edit from a Tor exit node IP within a period that the node was active. The IP addresses of users who are logged into accounts are not retained by the Wikimedia Foundation for more than a short period of time and are never made public. As a result, we could not identify edits made by registered Wikipedia users using Tor. 
Finally, we queried the timestamps of the identified revisions in the Tor relay search tool called ExoneraTor to verify that the IP addresses were indeed active exit nodes around the same time. We extracted and found a total of 11,363 edits on English Wikipedia made by Tor users between 2007 (the earliest available Tor consensus data) and March 2018 when our Wikipedia database dump was created. 

\figurename~\ref{fig:monthly-edits} displays the number of Tor edits to English Wikipedia per month over time. The spikes in the graph suggest that there were occasions when Wikipedia failed to ban exit nodes and Tor revisions were able to slip through. These larger spikes appear at least five times in the graph before late 2013, when the edit trend finally died down and failed to rise back up again.
We have posted the full dataset of Tor edits to Wikipedia and the code we used to conduct these analyses in a repository posted to the Harvard Dataverse where they will be available by request.\footnote{\url{https://doi.org/10.7910/DVN/O8RKO2}}

\subsection{How Wikipedia blocked Tor over time}

To better understand why Tor users were able to edit Wikipedia at certain times but not others, we examined the history of Wikipedia's Tor blocking and banning mechanisms. We found that there are two ways Wikipedia members prevent Tor users from editing: (1) \textit{blocking} the IP address using the TorBlock\footnote{\url{https://www.mediawiki.org/wiki/Extension:TorBlock} (Archive: \url{https://perma.cc/G44N-Y75R})} extension for MediaWiki, the software that was installed on the servers that run Wikipedia, and (2) \textit{banning} by blacklisting individual exit node IP addresses in a piece-meal process conducted by individual administrators and bots on Wikipedia. In 2008, Wikipedia started using the TorBlock extension to block Tor. TorBlock is a script that ``automatically applies restrictions to Tor exit node's access to the wiki's front-door server.'' This extension preemptively limits access from all active Tor nodes by pulling the current exit list published by Tor, as described in §\ref{sec:identifyingtor}. One benefit of using TorBlock is that only active Tor exit nodes are  prevented from creating accounts and editing. As soon as IP addresses stop volunteering as Tor exit nodes, they are restored to full access by TorBlock. However, as described by a Wikipedia administrator, the TorBlock extension did not seem to work well initially and also went down occasionally.\footnote{\url{https://en.wikipedia.org/wiki/Wikipedia:Bots/Requests_for_approval/TorNodeBot} (Archived: \url{https://perma.cc/SGS2-7BMZ})} As a result, Wikipedia administrators continued to issue bans manually and relied on bots to catch Tor nodes that were able to slip through.

Using publicly available data that Wikipedia maintains on bans, we traced the list of banned Tor IPs from 2007 to 2018. Wikipedia's block log provides details about the timestamp of each ban action, the enforcer's username, the duration of the ban, and optional comment justifying bans. Unsurprisingly, most IPs in this list are described as being banned simply because they are Tor exit nodes. 
Table \ref{tab:block-actions} provides an overview of the ban actions against Tor IP addresses over the course of 11 years. There were a total of 45,130 ban actions against IP addresses that were used as Tor exit nodes during this period. Roughly 11\% of these bans were against Tor IPs that successfully made at least one edit. 
Ban actions executed before a single edit took place suggest that many IP addresses were preemptively banned by Wikipedia. We found that less than 2\% of the ban actions explicitly state that they are due to vandalism. On the other hand, 77.1\% of the actions mention the word ``Tor.'' These statistics provide both a picture of Wikipedia's policy in relation to anonymity-seeking users and a validation of our methodology for identifying Tor edits.

\begin{table}[t] \centering 
  \caption{Ban Actions Against Tor Exit Nodes} 
\begin{tabular}{ll}
\hline
Ban actions & Number \\
\hline
Ban actions against all Tor exit nodes & 45,130 \\
Ban actions against Tor exit nodes with at least one edit & 4,964 \\
Number of Tor exit nodes banned & 32,947 \\
Number of Tor exit nodes with at least 1 edit banned & 2,148 \\
Ban actions citing vandalism & 532 \\
Ban actions citing Tor ban policy & 34,797\\
\hline
\end{tabular} 
\label{tab:block-actions}
\end{table} 

Bans on Wikipedia can be issued by either administrators or bots. Our data on ban actions shows that, initially, Tor IP bans were mainly handled by administrators with 95.9\% of 7,852 ban actions issued by administrators from 2007 to 2009. Bans during this period were typically 1--5 years in duration. However, IP addresses typically spend only a short period of time volunteering as Tor exit nodes.\footnote{\url{https://nymity.ch/sybilhunting/uptime-visualisation/} (Archived: \url{https://perma.cc/MH2P-CFWN})} Banning these IPs for extended periods of time prevented these addresses from editing on Wikipedia even when they were no longer Tor nodes. From 2010 to early 2014, Wikipedia started employing bots to automatically spot and blacklist Tor nodes. During this period, the typical ban duration was reduced to two weeks. Although many exit nodes were only active for a portion of this ban period, some large nodes were active for much longer. In some cases, bans expired while a node was still active and, as a result, we found many nodes were banned multiple times with multiple edits made between bans. 

Additionally, Tor users frequently slipped past Wikipedia's TorBlock in systematic ways that appear to explain the sharp drop in the number of Tor edits from 2007 to 2009 and frequent spikes in edits from 2010 to 2013. A Wikipedia administrator explained that the TorBlock tool only checked for the current list of Tor nodes, but when some of them were shut off abruptly, their server descriptors were no longer published on the exit list.\footnote{\url{https://en.wikipedia.org/wiki/Wikipedia:Bots/Requests_for_approval/TorNodeBot}  (Archived: \url{https://perma.cc/SGS2-7BMZ})} If the IP addresses were then reused as Tor nodes, they did not reappear on the list for some time and escaped the TorBlock extension's notice. As a result, the admin wrote an automated tool named TorNodeBot to spot and ban any Tor node with access to Wikipedia editing.\footnote{\url{https://en.wikipedia.org/wiki/User:TorNodeBot} (Archived: \url{https://perma.cc/VPM4-75PZ})} TorNodeBot was active from 2010 to 2014 and is recorded to have issued 32,123 bans on 21,837 different Tor IP addresses during this period.

The deactivation of TorNodeBot in early 2014, along with the significant drop of Tor edits and banning actions against Tor nodes, suggests that the TorBlock extension started working as intended at this point in time. Only 562 edits were made by Tor users after 2013. We suspect that these edits are allowed because TorBlock must periodically pull the currently active exit list from Tor, which leaves a time gap when freshly activated nodes are not caught by the tool.

\section{Statistical Comparison of Tor Edits To Other Groups of Users}
\label{sec:comparing}

In addition to our dataset of Tor edits, we developed datasets from three comparison groups---IP editors, First-time editors, and Registered editors. IP editors are not logged into an account so that their edits are credited to their actual (i.e., non-Tor) IP address. 
The second group includes registered editors making their first contribution. 
The third group includes registered users who have made more than one edit before the edit in question. 
For each of these populations, we cannot know if the people editing have other accounts or if they have contributed from other IP addresses.
We randomly picked the same number of revisions from each group, time-matched with the original dataset, by determining the number of edits made each month by Tor users and then randomly picking the same number of edits made by each comparison group within the same month. 


To assess the quality of contributions, we used several measures of quality that were developed within the Wikipedia community and by social computing researchers. 
Before examining the quality of these edits, however, it is important to note that not all Wikipedia pages serve the same purpose. Although article pages are the most visible, Wikipedia contains many other pages devoted to discussion, coordination, user profiles, policy, and more. While Wikipedia has strict guidelines about editing article pages, other types of pages tend to have more relaxed standards.\footnote{\url{https://en.wikipedia.org/wiki/Wikipedia:Namespace} (Archived: \url{https://perma.cc/P2ZP-R4TQ})} Although sections §\ref{sec:reverts} and the analysis of reverts in §\ref{sec:post2013} uses data drawn from contributions to all types of pages, the rest of our analysis is restricted to edits made to article pages (called ``namespace 0'' pages in Wikipedia). We focused our analysis on article pages for two reasons. First, article production is the primary work of the Wikipedia community, and contributions here have the potential to be of the greatest value. Second, the nature of article contributions lend themselves to large-scale computational analysis better than discussions about policy and social interactions that require substantial interpretation in order to be assessed for value. In addition, the current version of TorBlock (and other forms of blocks and bans used in the past) permit IP addresses to edit their own user talk pages in order to allow them to contact administrators and appeal their ban. These pages are therefore not included in our analyses. It is important to note that the distribution of edits across namespaces is different across the four comparison groups. For example, Tor editors make a larger proportion of contributions to article pages than Registered users. The distribution of edits across namespaces is available in the Appendix (\figurename~\ref{fig:np_distribution}).

Because the number of contributions to Wikipedia from Tor shrank drastically by the end of 2013, we divided and observed the edits in two separate periods from 2007 to 2013 and from 2013 to 2018. Because §\ref{sec:reverts} through §\ref{sec:tokens} are focused on identifying trends over time, we limit our analysis to the pre-2013 datasets where data is more dense. We replicated and compared results from §\ref{sec:reverts} through §\ref{sec:tokens} in the 2013--2018 data which we report on in §\ref{sec:post2013}. In all other sections, we conducted analyses using the full 2007--2018 dataset.

\subsection{Measuring contribution quality using reversion rates}
\label{sec:reverts}

The most widely used method for measuring edit quality in Wikipedia is whether an edit has been reverted. In Wikipedia, a contribution is said to be reverted if a subsequent edit returns a page to a state that is identical to a point in time prior to the edit in question and if the reverting edit is not reverted itself. Because the term ``revert'' can be used in a more general sense, these are sometimes called ``identity reverts.'' Because reverting is the main way that Wikipedia editors respond to low-quality contributions and vandalism \cite{Kittur:2007:HSS:1240624.1240698}, the reversion rate can provide insight into how valuable the efforts of an editor are perceived to be by the Wikipedia community. 

\begin{figure}
\centering
\includegraphics[width=\columnwidth]{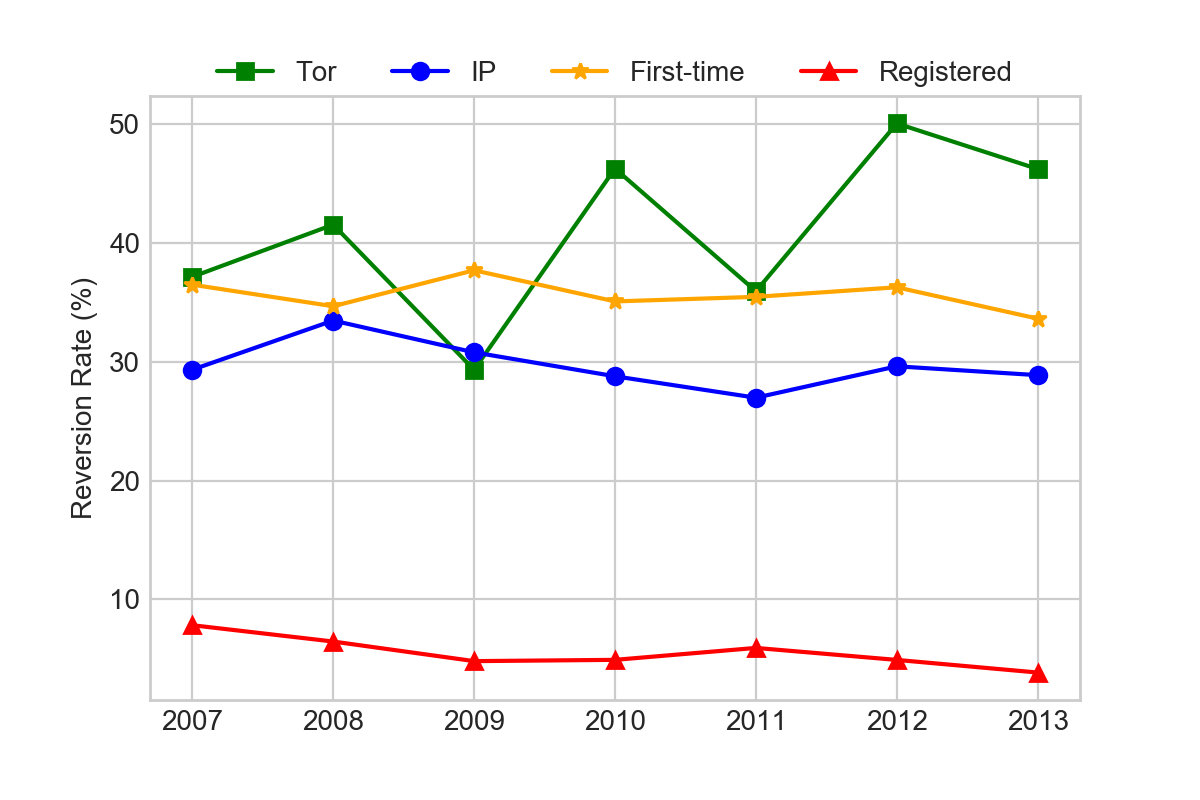}
\caption{Reversion rate for edits from different groups of editors over time (2008--2013).}
\label{fig:reversion-rates}
\end{figure} 

We used a Python library called \emph{mwreverts}\footnote{\url{https://pythonhosted.org/mwreverts/} (Archived: \url{https://perma.cc/HG6U-U5K2})} to detect whether or not a revision was subsequently reverted by someone else and whether or not an edit was was a revert action itself undoing other revisions. We examine the reversion rate of each set of edits in our comparison groups---both overall and by year. \figurename~\ref{fig:reversion-rates} plots how the reversion rate of article pages changes over time for each group of editors. Overall, 41.12\% of Tor edits on article pages are reverted, while only 30.3\% of IP edits, 35.2\% of First-time edits, and  5.5\% of Registered edits are reverted. A proportional z-test shows that the reversion rate of Tor edits is significantly higher than the closest group, First-time edits ($z = 11.53; p < 0.01$).\footnote{A Bonferonni correction for tests against our three comparisons groups results in an adjusted threshold of $\alpha=0.017$. 
We use this threshold when reporting statistical significance throughout. It is worth noting that because many of our findings are null results, an unadjusted $\alpha=0.05$ threshold is more conservative.} These numbers are similar for the reversion rates across all edits (special namespace articles included): 42.0\% for Tor edits, 29.61\% for IP edits, 34.3\% for First-time edits, and 4.84\% for Registered edits.

Reversion rate might be a biased measure of quality because good quality edits made via Tor might be reverted simply because they violate Wikipedia's policy blocking Tor. To assess whether this is in some cases true, one of the authors examined the 4,972 instances in which a Tor-based editor's work was reverted and hand coded the ``edit summaries'' left behind by the person performing the reversion. In 2,848 instances (57.3\% of the cases), no edit summary was entered as part of the revert action. Of the 2,124 reverts where the person doing the reverting provided an edit summary, 162 (7.6\% of reverts giving a reason) referred to conditions relevant to being a Tor user, with one or more of the following keywords: ``Tor,'' ``sock,'' ``block'', ``ban'' (referring to the ban policy of Tor IPs), ``proxy,'' ``masked,'' ``puppet,'' ``ip hopper,'' ``no edit history,'' ``multiple IP,'' ``dynamic IP,'' or ``log in'' (as in ``please log in'' or ``you can't log in'').
To the degree that other community members were more suspicious of Tor editors, reversion rate may be underestimating the quality of their contributions.


\subsection{Revert actions and their success rate}
\label{sec:revert_rates}

\begin{table*}
\centering
\begin{threeparttable}
\caption{Revert actions and Revert success rate}
\centering
 \begin{tabular}{l c l c l} 
 \hline
 Group & Revert actions\tnote{1} & Reverts kept\tnote{2} & Non-revert actions & Non-reverts kept\tnote{3}\\ [0.5ex] 
 \hline
 Tor editors  & 1,132 & 333 (29.41\%) & 6,619 &  4,224 (63.81\%) \\
 IP editors & 411 & 291 (70.80\%) & 10,040 & 7,117 (70.88\%)  \\
 First-time editors & 398 & 254 (63.81\%) & 7,878 & 5,095 (64.67\%) \\
 Registered editors & 1,189 & 1,049 (88.22\%) & 5,932 & 5,751 (96.94\%)  \\
 \hline
 \end{tabular}
\begin{tablenotes}
     \item[1] Revert actions: Edits that revert other edits.
     \item[2] Reverts kept: Revert actions that are not reverted by other edits.
     \item[3] Non-reverts kept: Edits that do not revert other edits and do not get reverted.
   \end{tablenotes}
 \label{tab:reversion-use}
 \end{threeparttable}
\end{table*}

A study of contributions to Wikipedia by Javanmardi et al.~\cite{javanmardi} showed that IP editors' contributions were twice as likely to be reverted and that registered users were almost three times more likely to revert another user as IP editors. We found that the latter is not the case for Tor editors. As illustrated in Table \ref{tab:reversion-use}, we found that Tor users, similar to Registered users, are much more likely than IP editors to revert others. Although Tor users are still statistically less likely to revert edits than Registered users ($z = -5.19, p < 0.01$), less than one third of their revert actions are allowed to stand by the Wikipedia community. This paints a stark contrast with the other groups whose revert actions are all much more likely to be kept. Overall, Tor editors revert others more frequently but less effectively. This points to an important difference in the behavior of Tor users and our comparison groups. When we excluded these cases of reverted reverts, Tor edits are much more likely to be kept. Indeed, non-reverts by Tor users are accepted at a rate that is comparable to First-time editors ($z = -1.44; p = 0.15$).

A deeper look into Tor revert actions reveals additional insights. First, Tor users are more likely to revert edits to non-articles. 28.2\% of Tor users' revert actions focus on non-article namespace articles while less than 12\% of revert actions from other groups do so. Tor users' reverts to non-articles are themselves reverted 85.16\% of the time. 
We also find that these revert actions primarily target Talk pages, such as Article Talk pages, and User Talk pages. 

Research by Yaserri et al.~has shown that a ``considerable portion of talk pages are dedicated to discussions about removed materials and controversial edits'' \cite{10.1371/journal.pone.0038869}. These discussions often resulted in extended back and forth between those editors who rarely change their opinion and can often lead to ``edit wars.'' An edit war happens when ``editors who disagree about the content of a page repeatedly override each other's contributions,'' changing the content of the page back and forth between versions~\cite{Buriol}.\footnote{\url{https://en.wikipedia.org/wiki/Wikipedia:Edit_warring} (Archived: \url{https://perma.cc/W5UZ-L4YD})} In November 2004, the Wikipedia community issued a guideline known as the three-revert rule (3RR), which prohibits an editor from performing ``more than three reverts, in whole or in part, whether involving the same or different material, on a single page within a 24-hour period.'' Anyone who violates this rule is at risk of being banned by Wikipedia administrators. In this way, the 3RR creates an incentive to seek anonymity.

To identify edit wars and violations of the 3RR, we examined the revision history of Tor edits in chronological order. We excluded self-revert actions because reverting one's own edit is allowed. Among 1,577 Tor revert actions, we found 30 3RR violations with a total of 180 revert actions made across 30 different articles. While the edit wars in our dataset rarely lasted more than several days and most of these violations did not last long before the Tor IP addresses were banned, this analysis provides evidence that Tor was used to engage in edit warring in violation of Wikipedia policy. We further reviewed these reverts and found that 56\% of the 180 edits are made on User Talk pages. A common pattern involved a Tor user reverting warning messages posted by Wikipedia administrators about vandalism. Unsurprisingly, 169 out of 180 Tor edits that were involved in edit wars were reverted as part of the back-and-forth conflict.

This is a conservative measure of edit warring by Tor users. Because of the dynamic nature of Tor IP addresses, Tor users can simply change to a different exit address to avoid being flagged by automated tools enforcing 3RR. As a result, we expanded our search to find any series of more than two reverts made on a single page within 24-hour period from \textit{any} Tor IP address. We found 546 total revisions, with 102 potential incidents in violation of the 3RR. Our manual inspection of dozens of these incidents suggests that, even when reverts are made from different Tor exit node IPs, pages were typically reverted to an older revision made by another Tor IP. This suggests it was the same person using different exit nodes making these reverts.
Once again, the chance of these reverts on article pages staying untouched was unlikely and 88.2\% of them were ultimately reverted. 

Because our Tor dataset includes the entire population of Tor edits, we could conduct an analysis of Tor being used to violate 3RR. Because our comparison sets are random samples, they are unlikely to contain consecutive edits made by the same user. To obtain some estimate of the rate at which other populations violate the 3RR, we retrieved all Wikipedia reverts made within the 48-hour period following each revert in all three of our comparison groups. 
Similar to findings in previous research, we found that other user groups are extremely unlikely to violate the 3RR policy \cite{10.1371/journal.pone.0038869}. In stark contrast to our Tor edits, we detected only 13 violations of the 3RR across all three comparison groups.
This relatively widespread rate of edit wars among Tor edits reflects the most important difference between Tor editors and our comparison groups identified in our analyses.

\subsection{Measuring contribution quality using persistent token revisions}
\label{sec:tokens}

 
Although an edit is only treated as an identity revert if it returns a page to a state that is identical to a previous state, contributions might also be removed through actions that add other content or change material. As a result, reverts should be understood as a particular and very conservative measure of low-quality editing. A more granular approach to measuring edit quality involves determining whether the parts of a contribution continue to be part of the article over multiple future revisions. 
According to Halfaker et al.~\cite{Halfaker:2009:JYP:1641309.1641332}, the survival of content over time can give important insights about a contribution's resistance to change and serves as a measure of both productivity (how much text was added) and quality (how much was retained) for a given revision.

Our approach used the \textit{mwpersistence}\footnote{\url{https://pythonhosted.org/mwpersistence/} (Archived: \url{https://perma.cc/P2F9-CA28})} library to calculate the number of words or fragments of markup (``tokens'') added to the articles in a given edit and then to measure how many of these tokens persist over a fixed window of subsequent edits. Following previous work, our measure of persistent token revisions (PTRs) involves collapsing sequential edits by individual users and then summing up every token added in a given revision that continues to persist across a window of seven revisions~\cite{Narayan:2017:WAF:2998181.2998307}. This measure only takes non-revert edits into account because revert actions always have 0 PTR.

\begin{figure*}[t]
\centering
\includegraphics[width=\textwidth]{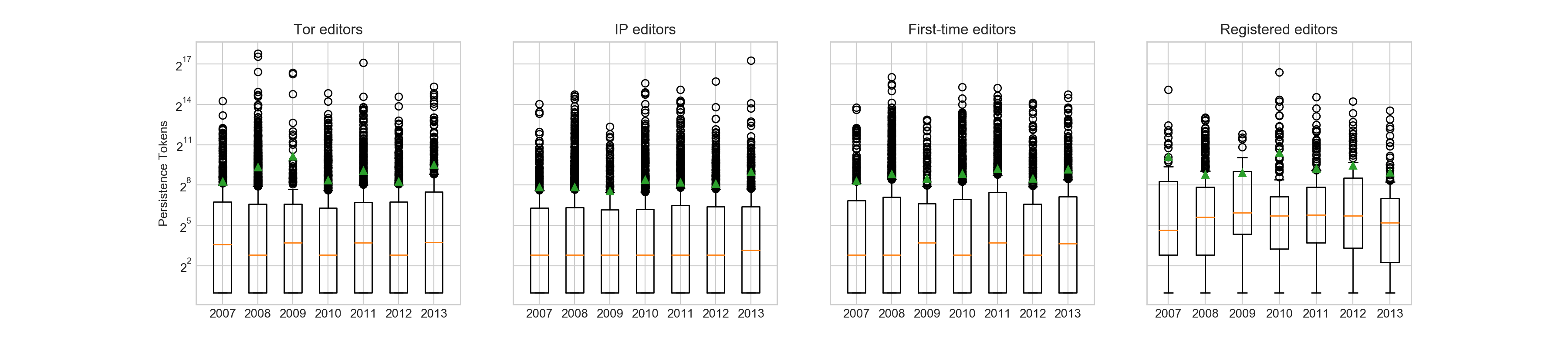}
\caption{Measurement of PTRs of different groups of non-revert edits over time. The rectangle is the interquartile region (middle 50\% of the population), with an orange line at the median. The upper and lower whisker represent the range of the population that is 1.5 times above and below the interquartile range. The green triangle is the mean, and the circles indicate individual observations falling outside the limit.}
\label{fig:persistence_boxplot}
\end{figure*}

\figurename~\ref{fig:persistence_boxplot} describes the contribution quality of non-revert revisions estimated by measuring PTRs for each edit between 2007 to 2013. We used a box plot to depict the distribution of PTRs for edits made in each year. Apart from Registered editors, the minimum value and the 25\% quartile of other groups are all 0. This reflects the fact that many edits to Wikipedia remove tokens instead of adding them and lead to a PTR count of 0. Edits that are entirely reverted also have a count of 0. The medians of the first three groups are relatively low, mostly within the range of 0 to 10 tokens. Registered editors' medians are higher, within the range of 10 to 40 tokens. The interquartile regions (IQRs) in the plots of Tor editors are slightly higher than those of IP editors and are comparable to those of First-time editors. The triangles on the graph display the mean PTR each year. Tor editors have some exceptional contributions outside the 95\% interval, which increases this mean value. Overall, we calculated the mean number of PTRs contributed by Tor editors as 547, by IP editors as 282, by First-time editors as 456, and by Registered editors as 836. Mann-Whitney U-tests suggest that Tor-edits have significantly higher PTRs than IP-edits ($U = 18158458, p < 0.01$) and First-time edits ($U = 14104155, p < 0.01$), but significantly lower than Registered edits ($U = 3095249, p < 0.01$). This provides evidence that contributions coming from Tor nodes have relatively significant value in terms of both quantity and quality as measured by PTRs.

\subsection{Analysis of reversion rate and persistent token revisions after 2013}
\label{sec:post2013}

As described in §\ref{sec:comparing}, Wikipedia's effort to block Tor users made it much harder for an edit to slip through by the end of 2013. In this section, we consider the small number of edits made in this later period. Using the methods described above, we computed the reversion rate and the PTRs for the population of 536 edits made after 2013 along with the same number of time-matched edits from other groups as described above. The results of this analysis are reported in Table \ref{tab:post2013}.

\begin{table*}[t]
\caption{Revert and PTR analyses of edits made after 2013}
\centering
 \begin{tabular}{l l l l l} 
\hline
 & Tor editors & IP editors & First-time editors & Registered editors \\ [0.5ex] 
 \hline
 Reversion rate & 28.2\%  & 25.0\% & 30.0\% & 5.7\% \\
 Revert actions & 38 (7.0\%) & 28 (5.2\%) & 10 (1.8\%) & 24 (4.5\%) \\
 Mean of PTRs & 645 & 162 & 310 & 3121  \\
 Median of PTRs & 12 & 6 & 0 & 18 \\
 \hline
 \end{tabular}
 \label{tab:post2013}
\end{table*}

Compared to the number we see from the 2007--2013 period, Tor's reversion rate decreased from 42.1\% in the period before December 2013 to 28.2\% afterward. Two other comparison groups (IP editors and First-time editors) also exhibit a decline in the rate of edits being reverted. This reflects the fact that reversion rates have been in decline in Wikipedia over time in general.\footnote{\url{https://stats.wikimedia.org/EN/EditsRevertsEN.htm} (Archived: \url{https://perma.cc/7WY8-MS6P})} 
Due to the small number of edits each year, we were unable to properly observe whether the change happened gradually or as a result of Wikipedia's more effective quality-checking methods. 
Overall, the reversion rates of Tor editors are now statistically comparable to IP editors ($z= 0.89; p = 0.19$), and First-time editors ($z= -0.67; p = 0.25$).
In terms of revert actions, we see a significant decline in the number of revert actions that Tor editors took ($z=-2.5; p < 0.01$) as well as in all our comparison groups. Overall, Tor editors' revert rate in the later period are comparable to that of IP editors ($z= 1.2; p = 0.10$) and that of Registered editors ($z= 1.83; p = 0.03$), but still higher than that of First-time editors ($z= 2.97; p < 0.01$).

Our measure of PTR also suggests that Tor editors are at least as high quality as IP editors and First-time editors in the post-2013 period. Mann-Whitney U tests suggests that Tor edits made after 2013 are of similar quality to edits by IP editors ($U=48142; p=0.118$), of greater quality than edits by First-time editors ($U=49692; p<0.01$), but are of lower quality than those by Registered editors ($U=9684; p=0.02$).  This final difference is not statistically significant after a Bonferroni adjustment for multiple comparisons.
While it is clear that contributions from Tor users significantly improve in many aspects after 2013, we also observe a similar pattern in IP editors and First-time editors. As a result, it is hard to argue that the increasing effectiveness of TorBlock extension is the sole reason for this change.

\subsection{Measuring quality through manual labelling}
\label{sec:handcode}

Perhaps the most compelling way to assess the quality of Tor edits is to categorize edits manually. To do so, we conducted a formal content analysis of edits. Two of the authors and two colleagues conducted a content analysis following guidelines laid out by Neuendorf \cite{neuendorf_content_2017} to code revisions as Damaging or Non-Damaging. To ensure that we had a large enough sample, we first conducted a simulation-based power analysis, which indicated that a sample of 850 edits in each group would be necessary to detect an underlying difference of 7\% in the proportion of damaging edits between groups at the $\alpha=0.05$ confidence level.\footnote{A power analysis requires a minimum effect size, and we chose 7\%.} The team developed a codebook, and after conducting three rounds of independent coding followed by discussion of codes to develop a shared understanding and definitions, we drew a year-matched random subsample of 999 edits from our sample of Tor edits and the three comparison datasets.

 We defined damaging edits as those we would want to remove from the encyclopedia because they diminished the usefulness of the resource by being incorrect, sloppy, a violation of Wikipedia style, or by otherwise causing the article to be less encyclopedic. Some edits were observed to contain both mistakes and positive contributions. We used our judgment to assess whether the contribution was generally positive and worthwhile, despite being imperfect. 
When we did not see evidence that led us to suspect that an edit was damaging, we followed Wikipedia's convention of assuming good faith and coded it as Non-Damaging. 

Edits were presented to coders as a ``diff'' that showed what was changed using the same interface that Wikipedia contributors can use to review contributions and were presented in a randomized order using filtering software to suppress identity information about contributors. 
Coding was conducted without reference to other contextual information, including subsequent or previous edits.
Four coders conducted independent coding and discussion of codes over several rounds. Subsequently, they classified a dataset of 160 edits (40 from each group) and compared their results (10 assessments were missing from one coder). This result was a good level of inter-rater reliability across the four coders (raw agreement of $89\%$; pairwise agreement of $80$\%; Gwet's AC of $0.68$).\footnote{Gwet's AC was used because it is a measure of multi-rater reliability robust to variation in the distribution of units that raters encounter \citep{quarfoot_how_2016}.} 
Full agreement is unlikely because our protocol required coders to rely on their judgement and knowledge to detect things like misinformation without recourse to any outside information.
The full hand-coded sample includes the consensus rating of the 160 edits evaluated in the pilot plus 800 random edits drawn from subsamples described earlier that were coded by each of three researchers and 840 edits coded by the fourth. 
We omitted 30 revisions from our final analysis because they were missing or otherwise deleted from Wikipedia. 

\begin{table}
\centering 
  \caption{Results from logistic regressions of hand-coded quality assessments of edits. Tor editors served as the omitted category.} 
  \label{tab:logitHand} 
\begin{tabular}{l c }
\hline
 & Non-Damaging \\
\hline
Intercept                                  & $0.85^{*}$        \\
                                             & $[0.71;\ 1.00]$   \\
First-time Editors & $-0.25^{*}$       \\
                                             & $[-0.46;\ -0.05]$ \\
IP-based Editors    & $0.10$            \\
                                             & $[-0.12;\ 0.31]$  \\
Registered Editors   & $1.69^{*}$        \\
                                             & $[1.39;\ 1.99]$   \\
\hline
AIC                                          & 3551.08           \\
BIC                                          & 3575.53           \\
Log Likelihood                               & -1771.54          \\
Deviance                                     & 3543.08           \\
Num. obs.                                    & 3337              \\
\hline
\multicolumn{2}{l}{\scriptsize{$^*$ indicates that 0 is outside the 95\% confidence interval.}}
\end{tabular}
\end{table}

The results of the from logistic regression using Tor-based edits as the baseline are reported in Table \ref{tab:logitHand}.
We found that 70.1\% of edits made by Tor-based editors were coded as Non-Damaging, while 72.1\% of edits by IP-based editors and 64.6\% of edits by First-time editors were. Although slightly higher and lower respectively, our model suggests that the proportion of Non-Damaging edits was not statistically different than our sample of Tor edits in these two comparison groups. We found that 92.7\% of edits by Registered editors were Non-Damaging---a statistically significant difference from our sample of Tor edits.

\section{Classification of Edits using Machine Learning Tools}
\subsection{Measuring contribution quality using ORES}
\label{sec:ores_long}

Wikimedia uses a machine learning system called ORES to automatically categorize the quality of contributions to Wikipedia \cite{halfaker_ores:_2016}. The system was developed to support Wikipedia editors trying to protect the encyclopedia from vandalism and other kinds of damage. With assistance from the ORES team, we used the system to assess the quality of the edits in our comparison groups. Because ORES is fully automated, we were able to conduct our analysis on the full datasets. ORES classifies edits in terms of the likelihood that they are ``Good Faith'' and ``Damaging''~\cite{halfaker_ores:_2016}. We recoded Damaging as Non-Damaging so that in all cases ``high'' scores are positive and ``low'' scores are negative.

While there exists no gold standard set of features for assessing the quality of work on Wikipedia \cite{7809715}, ORES is trained using edit quality judgments solicited from the Wikipedia community. The system uses 24 different features for English Wikipedia \cite{Dang:2016:QAW:2996442.2996447, Warncke-Wang:2015:SFQ:2675133.2675241, Warncke-Wang:2013:TMM:2491055.2491063}. These include the presence of ``bad words,'' informal language, whether words appear in a dictionary, repeated characters, white space, uppercase letters, and so on. Other features are related to the amount of text, references, and external links added or removed in a revision. In addition to features related to the text of a contribution, ORES uses contribution metadata such as whether the editor supplied an edit summary, and contributor metadata such as whether the editor is an administrator or is using a newly created account. The specific list of features differs by language, and a full list is available in the publicly available ORES source code.\footnote{\url{https://github.com/wikimedia/editquality/tree/master/editquality/feature_lists} (Archived: \url{https://perma.cc/TME4-NSL6})} Previous work has found that ORES scores are systematically biased so that it classifies edits by IP editors and inexperienced users as being lower quality~\cite{halfaker_ores:_2016}. 

\begin{figure}[t]
\centering
\includegraphics[width=\columnwidth]{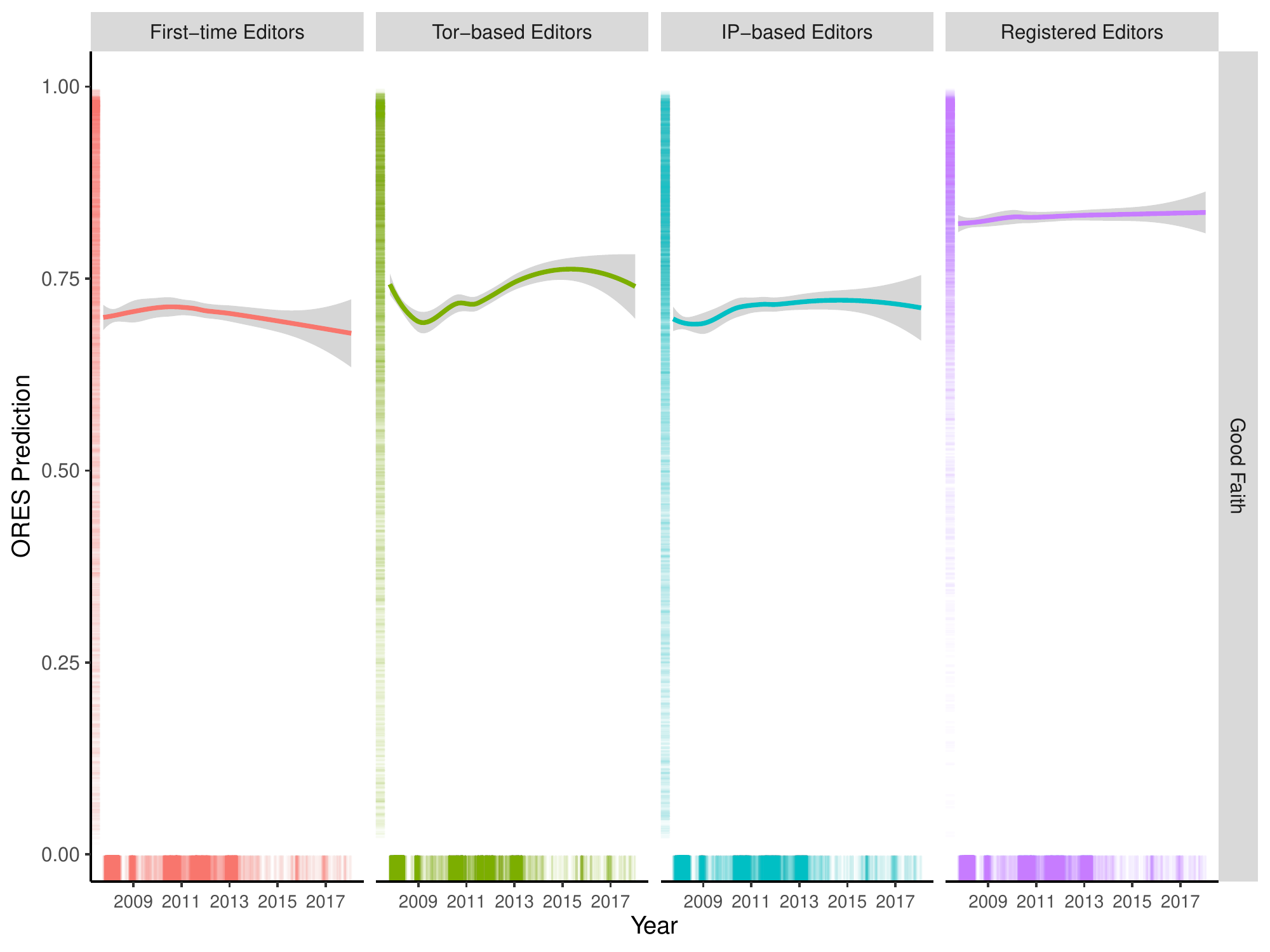}
\caption{A non-parametric LOESS curve over time. We use feature injection to instruct the ORES Good Faith model to treat all edits as if they were made by a newly created user account.}
\label{fig:feat_loess} 
\end{figure}

\begin{table}
\centering
\caption{Logistic regression using a feature-injected ORES model. 
First-time editors served as the omitted category.
}
\label{tab:feat_ns0}

\begin{tabular}{l c c }
\hline
 & Good Faith & Non-Damaging \\
\hline
Intercept          & $0.87^{*}$       & $0.27^{*}$      \\
                   & $[0.82;\ 0.92]$  & $[0.22;\ 0.31]$ \\
Tor-based Editors  & $0.10^{*}$       & $0.14^{*}$      \\
                   & $[0.03;\ 0.17]$  & $[0.07;\ 0.20]$ \\
IP-based Editors   & $0.01$           & $0.07^{*}$      \\
                   & $[-0.06;\ 0.08]$ & $[0.01;\ 0.14]$ \\
Registered Editors & $0.70^{*}$       & $0.68^{*}$      \\
                   & $[0.62;\ 0.79]$  & $[0.61;\ 0.76]$ \\
\hline
AIC                & 26819.97         & 35414.08        \\
BIC                & 26853.08         & 35447.18        \\
Log Likelihood     & -13405.98        & -17703.04       \\
Deviance           & 7541.53          & 7395.66         \\
Num. obs.          & 29057            & 29059           \\\hline
\multicolumn{3}{l}{\scriptsize{$^*$ indicates that 0 is outside the 95\% confidence interval}}
\end{tabular}

\end{table}

To understand contribution quality independent of identity-based features, we made use of the ``feature injection'' functionality in ORES \cite{halfaker_ores:_2016}. Using feature injection, we instructed ORES to treat all revisions as if made by Registered users whose accounts are 0 seconds old. A visualization of the feature-injected ORES analysis of our comparison sets are shown over time in \figurename~\ref{fig:feat_loess}. This visualization is produced using LOESS smoothers \cite{cleveland_robust_1979}.\footnote{LOESS plots are a visualization tool that use low-order polynomial regression on each datapoint to calculate a smoothed fit line that describes the data as a weighted moving average. The grey bands represent standard errors around the LOESS estimates.} This model is of Good Faith measure; we omit the Non-Damaging ORES model because the lines are extremely similar. This visualization shows that Tor, IP, and First-time editors are all comparable, with Tor editors appearing to make slightly higher quality contributions than First-time and IP editors, particularly in the latter parts of the data. We used logistic regression to test for statistical differences, treating First-time editors as the baseline category as they most closely resemble our feature injection scenario. The results of our model are reported in Tab.~\ref{tab:feat_ns0}. 

The positive coefficient for Tor in both Good Faith and Non-Damaging scenarios indicates that Tor users are slightly better contributors than our baseline of First-time editors by the ORES measurement. Although the 
differences are statistically significant, 
the estimated chance that a given edit will be Good Faith at the baseline (new account) is 70.5\%. 
whereas the likelihood that an edit will be Good Faith if it originates from a Tor editor is 72.5\%. 
We believe that the estimated 2\% margin is unlikely to be practically significant. 
For the Non-Damaging model, we likewise find statistically significant differences between Tor edits and our comparison groups but also find that the practical effects are small. Our models predict higher average rates of Non-Damaging edits for Tor editors (60.1\% for Tor editors versus 56.7\% for First-time editors) and IP editors (58.4\%). For both models, contributions from Registered editors are estimated to be of high quality, with a prediction of 82.8\% Good Faith and 72.1\% Non-Damaging.
These results provide additional evidence in support of our hypothesis that Tor editors, IP editors, and First-time editors are quite similar in their overall behavior but that quality levels of contributions from Registered editors are higher. 

\subsection{Comparison of Hand-coded Results to ORES Results} 
\label{sec:hand_ores}
Given that we performed two different kinds of analysis to identify Non-Damaging edits (i.e., hand-coding the edits, and scoring via the ORES machine learning platform), we can examine the extent to which these two measures agree. Doing so is valuable because it can indicate whether the ORES classifications used by Wikipedia are systematically biased against contributors from Tor editors. 
As with our analysis in §\ref{sec:ores_long}, we used feature injection to instruct ORES to treat all edits in the hand-coded sample used in §\ref{sec:handcode} as if they were being made by newly Registered editors. We then used these data to compare the ORES prediction with and without feature injection to our manual assessment for all four user groups by generating receiver operating characteristic (ROC) curves. We have included the full curves in our appendix in \figurename~\ref{fig:ROC_nd}.

Table \ref{tab:algBias} reports model performance in the form of area under the curves (AUC) for the ROC curves for each of our comparison groups. These results indicate that there is substantial room for improvement in ORES. Using feature injection, ORES performs best relative to our hand-coded data when predicting the quality of edits performed by IP editors ($AUC = 0.811$ for Non-Damaging), less well for Tor editors ($AUC = 0.758$), and even less well for First-time editors ($AUC = 0.704$) but, strikingly, worst for Registered editors ($AUC = 0.663$).

When we examined a small sample of edits where our hand-coding and ORES disagreed, we found there were often good reasons for the disagreement. Our hand-coding process included doing work that ORES does not do, such as noticing when links were to personal or spam websites and weighing the context of the edit on the page against our own understanding of appropriate and correct encyclopedic content. These results suggest that machine learning tools such as ORES have a limited ability to assess the quality of edits without human intervention.


\begin{table}
\caption{Classifier AUC of ORES with and without feature injection for our four samples of edits.}
\centering
 \begin{tabular}{c c c} 
 \hline
 & AUC w/ Injection & AUC w/o Injection \\
 \hline
 First-time Editors & 0.704 & 0.708 \\
 IP Editors & 0.811 & 0.814 \\
 Tor Editors & 0.758 & 0.753 \\
 Registered Editors & 0.663 & 0.673 \\
\hline 
 \end{tabular}
 \label{tab:algBias}
\end{table}

Systematic bias in ORES could result in higher rates of rejection of contributions from some groups of editors. Feature injection as we have done it treats registered editors as if they are new---essentially removing a ``benefit of the doubt'' based on their longevity in the community. 
Table \ref{tab:algBias} shows that feature injection has very modest effects on model performance---dropping AUC by 0.01 for Registered editors and by 0.004 for First-time editors while improving AUC by 0.005 for Tor editors and by 0.003 for IP editors.


The team that developed ORES published a set of recommended operating points. For example, they suggest that users developing fully automated systems (``bots'' below) maximize recall at a precision of $\geq$90\%.
They suggest that users developing a human-involved system maximize filter rate (that is, the number that are not routed for review) at recall $\geq$75\%. ORES provides an interface to use preferred constraints to select an optimized decision-making threshold. For example, if we use the provided ``bot'' constraint, ORES recommends an operating point threshold of .055; that is a bot should only automatically discard an edit if the Non-Damaging level is below 5.5\%.  We examine our results using these thresholds to understand how ORES would classify Tor edits in Wikipedia's normal workflow.

\begin{table*}
\caption{Comparison of ORES developer-predicted performance to actual performance of our hand-coded sample of edits made from Tor without feature injection ($n=847$).}
\centering
 \begin{tabular}{p{1.2cm}p{1.5cm}p{1.6cm}p{1.3cm}p{1.3cm}p{1.3cm}p{1.3cm}p{3.0cm}}
 \hline
 Scenario & Optimizing Constraint & Recommended Threshold & Actual (Predicted) Accuracy & Actual (Predicted) Precision & Actual (Predicted) Filter Rate & Actual (Predicted) Recall & Result of Filtering \\
 \hline
Automatic Removal & Max. Recall at Precision~$\geq$~90\% & \textless5.5\% non~damaging & .713 (.913) & 1 (.909) & .988 (.998) & .040 (.045) & 10 of 847 dropped due to high confidence of damage; prior hand coding found all 10 of these to be damaging\\
Route for Human Review & Max. Filter Rate at Recall $\geq$ 75\% & \textless 68.6\% non~damaging & .574 (.904) & .396 (.226) & .386 (.887) & .814 (.751) & 520 of 847 routed for human review due to modest confidence of damage; prior hand coding found 206 of these routed edits to be damaging.\\
 \hline
 \end{tabular}
 \label{tab:noinj_thresh}
\end{table*}

The predicted values we report in Table \ref{tab:noinj_thresh} describe ORES' predictions about its own performance based on its training data using these recommended thresholds. Our results indicate that while a system that uses bots can identify a small proportion of damaging edits made through Tor, many damaging edits are missed while many Non-Damaging edits are routed for review. Our results suggest that ORES offers only moderate assistance to human-augmented systems seeking to review edits made by privacy seekers using Tor.

\subsection{Topic Modeling}
\label{sec:lda}

Although average quality may be similar, Tor editors may differ systematically from other editors in terms of what they choose to edit. Knowing which topics Tor users edit might provide insight into their reasons for seeking anonymity and the value of their contributions. For example, Tor users might pay more attention to matters that are sensitive and controversial. Unfortunately, the Wikipedia category system is an incredibly granular human-curated graph that is poorly suited to the construction of coarse comparisons across broad selections of articles \cite{thornton_tagging_2012}.

Topic modeling may assist such an exploration by offering clusters of keywords that can be interpreted as semantic topics present in a collection of documents. One of the most popular topic modeling techniques is called \emph{Latent Dirichlet Allocation} (LDA)---a generative probabilistic model for collections of discrete data such as text corpora \cite{blei2003latent}. \emph{Machine Learning for Language Toolkit} (MALLET) provides a widely used way to use LDA \cite{McCallumMALLET}. Given a list of documents and a number of topics, MALLET estimates a set of probability distributions of topics over the vocabulary of unique words. With these probability distributions and a further inspection of the keywords MALLET outputs, we can gain insight into the kinds of subjects that Tor users and other groups of editors pay attention to. While topic models are known to be unstable, they are useful for comparing documents across a set of \textit{ex ante} groups.

\begin{figure}[t]
\centering
\includegraphics[width=\columnwidth]{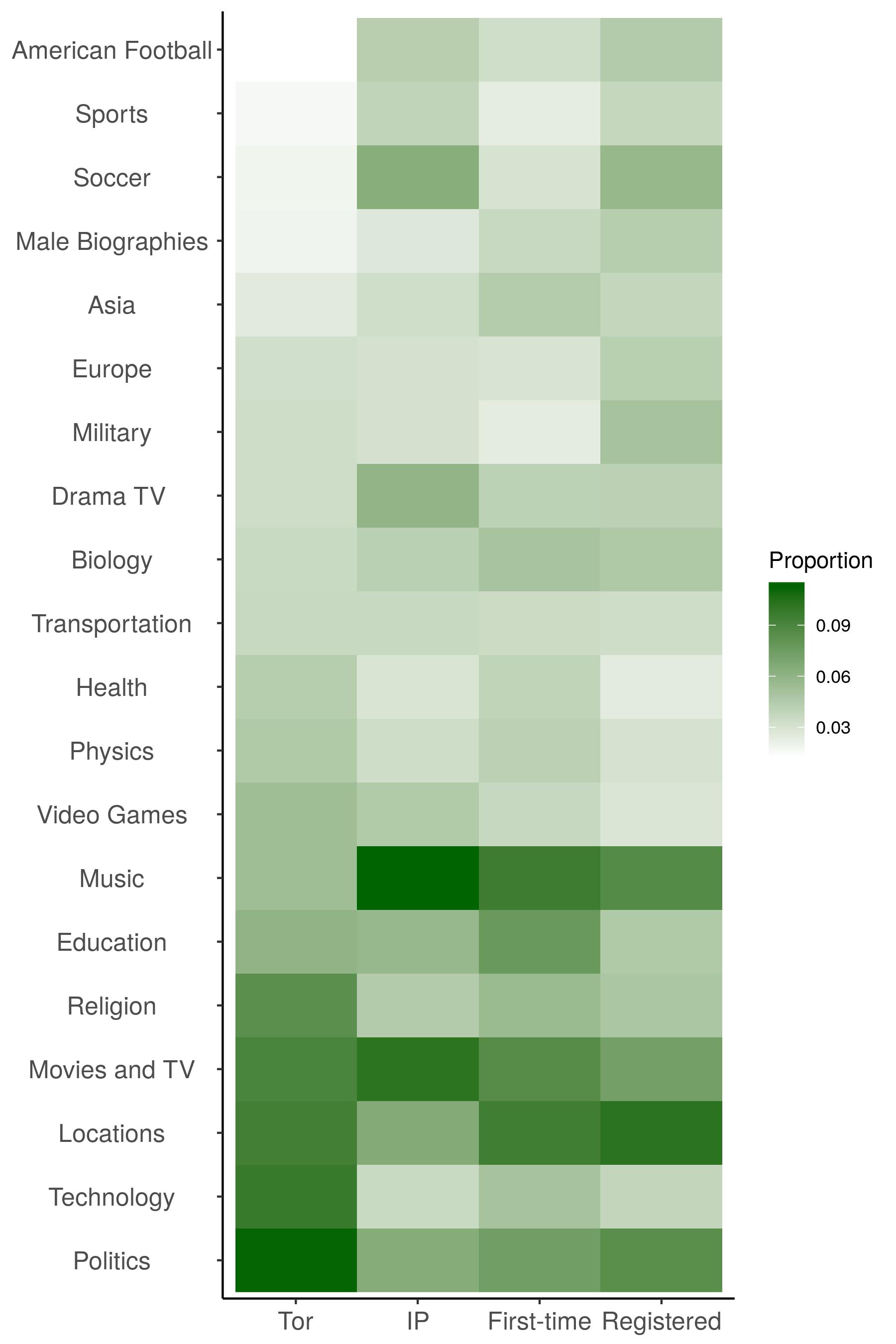}
\caption{A raster diagram showing the proportion of articles edited by each comparison group (along the $x$-axis) with where the topic (along the $y$-axis) is the single highest proportion.}
\label{fig:raster}
\end{figure}

Using our datasets of edits, we identified all the articles edited by Tor users and our three comparison groups. Next, we mined all textual content of these articles and then processed them through MALLET to produce keywords and their probability distributions. Because there is no optimal number of topics, we ran the tool to find 10, 20, 30, and 40 topics. For each number, we conducted four different runs to test the consistency of the results. After these experiments, we found that the results across the top five most frequent topics for each group of edits are highly consistent, with only slight changes in the keywords and the ranking. Because we felt that having 20 different clusters of keywords for the whole text corpora led to the most reasonable and comprehensible topics, the results reported below are from from LDA topic models estimated using 20 topics. All other parameters needed for the LDA algorithm were run with default values in MALLET. After fitting LDA topics models with MALLET, we manually interpreted each cluster of words and created an appropriate topic header. For reference, we include the mapping of keyword collections to topic headers we assigned in Table \ref{tab:mallet-topics} in our appendix. 

As a mixture model, LDA treats every document as belonging to every topic, but to varying degrees. As a result, we identified the topic with the highest probability and described each article as being ``in'' that topic for the purposes of the comparisons between the groups of edits. 
A Pearson's Chi-squared test suggests that the distribution of articles across topics is different between Tor editors and IP editors ($\chi^2=1655; df=19; p<0.01$), First-time editors ($\chi^2=848; df=19; p<0.01$), and Registered editors ($\chi^2=1508; df=19; p<0.01$). These differences are statistically significant after adjusting for multiple comparisons using a Bonferroni correction and suggest that Tor editors, although distinct from other groups of editors, are most similar to First-time editors in their topic selections. 

\begin{table}
\caption{Top 5 topics for each dataset}
\centering
 \begin{tabular}{c c c c} 
\hline
 Tor & IP & First-time & Registered \\ [0.5ex] 
 \hline
 Politics & Music & Music &  Locations \\ 
 Technology & Movies \& TV & Locations & Music \\
  Locations &  Locations & Movies \& TV & Politics \\
 Movies \& TV & Politics & Education &  Movies \& TV \\
 Religion & Sports & Politics & Sports \\
 \hline
 \end{tabular}
 \label{tab:top5topics}
\end{table}

Our analysis shows some similarities between Tor editors' interests and other groups. Table \ref{tab:top5topics} compares the top 5 topics that each group focused most on. \figurename~\ref{fig:raster} visualizes the distribution of topics using a gradient where more prevalent topics are darker and less prevalent topics are lighter. While there are many horizontal bands of a similar shade where the topics edited by our different sets of users are similar, we can also see many differences. 

For example, like other editors, Tor editors frequently edit topics such as \emph{Movies and TV} and \emph{Locations}, which are popular across all groups. 
We see proportionally fewer contributions from Tor editors in the \emph{Sports, Soccer}, and \emph{American Football} topics. 
Compared with other kinds of users, Tor editors are more likely to contribute to articles corresponding to \emph{Politics}, \emph{Technology}, and \emph{Religion}---topics that may be construed as controversial.\footnote{\url{https://www.thebalancecareers.com/topics-to-avoid-discussing-at-work-526267} (Archived: \url{https://perma.cc/G4GT-GEAK})} Our findings provide evidence to support previous qualitative work that has suggested that sensitive or stigmatized topics might attract Wikipedia editors interested in using tools like Tor to conceal their identity~\cite{Forte2017}.

\section{Limitations}

Our work is limited in several important ways.
First, our results are limited in that our analysis is conducted only on English Wikipedia. We cannot know how this work would extend to users of privacy-enhancing technologies other than Tor or to user-generated content sites beyond English Wikipedia.  As a minimal first step, we attempted to speak to this limitation by conducting an analysis of editing activity made by Tor users in other language editions of Wikipedia. Although we do not report on them in depth, we have included information in the appendix (see Tab.~\ref{tab:language-differences}) that displays the number of Tor edits in different language editions of Wikipedia relative to contributions made by the communities as a whole. Although Tor users are active in many language editions of Wikipedia, only a small number of edits by Tor users evaded the ban. 

There are reasons to imagine that the behavior of Tor editors contributing to English Wikipedia might differ from that of editors in language editions. For example, we identify thousands of edits from Tor exit nodes contributing to the Russian Wikipedia edition. This is striking because the Russian government partially bans access to Tor\footnote{\url{https://www.infosecurity-magazine.com/news/russia-passes-bill-banning-tor-vpns/} (Archived: \url{https://perma.cc/DLN7-KTQT})} and Wikipedia.\footnote{\url{https://en.wikipedia.org/wiki/Censorship_of_Wikipedia#Russia} (Archived: \url{https://perma.cc/GNM4-9UNH})} 

Although a closer inspection of Wikipedia language editions may yield interesting motivational and cultural differences regarding anonymity-seeking practice, our team is not sufficiently versed in these languages to conduct a replication of our analyses across different Wikipedia language editions. We are making our full datasets available and invite other researchers' interest.

Of course, Wikipedia language editions do not necessarily imply the geographic locations of editors. We do not know if people editing Russian Wikipedia come from Russia. Additionally, in many countries, viewers primarily access English Wikipedia even when English is not their native language.\footnote{\url{https://stats.wikimedia.org/wikimedia/animations/wivivi/wivivi.html} (Archived: \url{https://perma.cc/PGV6-687Q})} For example, the majority of pageviews from China and Iran---countries that ban both access to Tor and Wikipedia---go to the English version of Wikipedia. 
English Wikipedia is also the primarily-viewed Wikipedia for many countries that do not have a history of banning access to Wikipedia, such as the Netherlands and Croatia. 


Our study is limited in other ways as well. Because our study uses IP addresses and account names to identify editors, we cannot know exactly how usernames and IP addresses map onto people. Some users may choose different levels of identifiability depending on the kinds of edits they wish to make. For example, a registered editor may use Tor for certain activities and not for others \cite{Forte2017}. 

Additionally, our samples might reflect survivorship bias. We simply cannot know if our sample of Tor edits is representative of the edits that would occur if Wikipedia did not block anonymity-seeking users. Many Tor users who are told by Wikipedia that Tor is blocked will not try again. As a result, our dataset might overrepresent casual one-off Wikipedia contributors, including both constructive ``wiki gnomes'' and drive-by vandals. Our sample might also over-represent individuals with a deep commitment to editing Wikipedia or with technical sophistication (i.e., the knowledge that one could repeatedly request new Tor circuits to find exit nodes that are not banned by Wikipedia). Tor users who manage to evade the ban might include committed activists as well as banned Wikipedia users with deeply held grudges.
Although we do not know what \textit{else} would happen if Wikipedia unblocked Tor, we know that the almost total end of contributions to Wikipedia from Tor in 2013 means that, at a minimum, a large number of high-quality contributions are not occurring. Our analysis describes some part of what is being lost today---both good and bad---due to Wikipedia's decision to continue blocking users of anonymity-protecting proxies.

\section{Conclusions and Implications for Design}

Wikipedia's imperfect blocking of Tor provides a unique opportunity to gain insight into what might not be happening when user-generated content sites block participation by anonymity-seeking users. We employed multiple methods to compare Tor contributions to a number of comparison groups. Our findings suggest that privacy seekers' contributions are more often than not comparable to those of IP editors and First-time editors in many ways. Using hand-coded data and a machine-learning classifier, we estimated that edits from Tor users are of similar quality to those by IP editors and First-time editors. 
We estimated that Tor users make more higher quality contributions than other IP editors, on average, as measured by PTRs. 
Our analysis also pointed to several important differences.  We found that Tor users are significantly more likely than other users to revert someone else's work and appear more likely to violate Wikipedia's policy against back-and-forth edit wars, especially on discussion pages.
Tor users also edit topics that are systematically different from other groups. We found that Tor editors focused more on topics related to religion, technology, and politics and less on topics related to sports and music. 

The Tor network is steadily growing, with approximately two million active users at the time of writing. Many communities around the world face Internet censorship and authoritarian surveillance. In order to be Wikipedia contributors, these communities must rely on anonymity-protecting tools like Tor. In our opinion, our results show that the potential value to be gained by creating a pathway for Tor contributors may exceed the potential harm. 
Wikipedia's systemic block of Tor editors remains controversial within the Wikipedia community. We have been in close contact with Wikipedia contributors and staff at the Wikimedia Foundation as we conducted this research to ensure that our use of Wikipedia metrics is appropriate and to give them advance notice of our results. We are hopeful that our work can inform the community and encourage them to explore mechanisms by which Tor users might legitimately contribute to Wikipedia---perhaps with additional safeguards. Given the advances of the privacy research community (including anonymous blacklisting tools such as Nymble~\cite{nymble}), and improvements in automated damage-detecting tools in Wikipedia, alternatives to an outright ban on Tor contributions may be feasible without substantially increasing the burden already borne by the vandal-fighting efforts of the Wikipedia community. We hope our findings will inform progress toward these ends.

\section*{Acknowledgements}

We owe a particular debt of gratitude to Nora McDonald and Erica Racine who both contributed enormously to the content analysis included in the paper. Our methodology was improved via generous feedback from members of the Tor Metrics team, including Karsten Loesing, and the Wikimedia Foundation, including Aaron Halfaker, Morten Warncke-Wang, and Leila Zia.
Feedback and support for this work came from members of the Community Data Science Collective, and the manuscript benefited from excellent feedback from several anonymous referees at IEEE S\&P. The creation of dataset was aided by the use of advanced computational, storage, and networking infrastructure provided by the Hyak supercomputer system at the University of Washington.
This work was supported by the National Science Foundation (awards CNS-1703736 and CNS-1703049) and included the work of two undergraduates supported through an NSF REU supplement.

\bibliography{rachel}
\bibliographystyle{plain}

\appendix

\newpage
\begin{table*}[h!]
  \centering 
  \caption{Topic labels and clusters of keywords.} 
\begin{tabular}{l|l}
Topics & Keywords \\
\hline
 Soccer & league cup goals club team season stadium football match world\\ &clubs years united won final goal scored played caps win \\ 
  \hline
 Sports & score team world match championship win open round title seed\\ & won wrestling event time champion final defeated mexico san lost \\
  \hline
 Biology & species food water found made large sea fish animals common\\ & red island white small called years north animal south long 
   \\
  \hline
 Drama TV & back time family episode father death life man mother house \\ & season series make home son end find friend friends story \\
  \hline
 Military & war army military forces force battle british general air killed \\ & ship attack united u.s states troops police german soviet command \\
  \hline
  Locations & city area county park north river south west town station \\ & street population london state east district located road built national \\
  \hline
 Male Biographies & american john born james william george robert actor english player \\ & david british united york thomas michael henry charles years richard  \\ 
  \hline
 Health & health disease people treatment medical found research study sexual human \\ & blood risk effects cells children studies include symptoms brain age \\
  \hline
 Music & album music song band released single songs tour rock chart \\ & albums number records live guitar video year top label love \\
  \hline
 Technology & utc system data software windows talk users internet support information \\ & version wikipedia computer network systems page mobile user content web \\ 
  \hline
 Physics & energy water system light power time form space surface high \\ & called number process heat large field mass theory temperature gas  \\
  \hline
 Transportation & air airport aircraft company car engine international flight airlines system \\ & service cars speed model year production line design vehicles vehicle \\
  \hline
 American Football & season game team games player football league record teams year  \\ & won coach played bowl players win championship points nfl career \\ 
  \hline
 Religion & church book century god work life world early press history \\ & published society religious time people books modern jewish women christian  \\
  \hline
 Movies & film series show television award season episode awards role films \\ & episodes movie year september time production released actor comedy channel  \\
  \hline
 Video Games & game games series released character comics characters japanese player japan \\ & world version time video players story battle team original unknown  \\
 \hline
 Europe & french france century german russian empire europe european roman republic  \\ & population italian language greek king germany italy russia world spanish  \\
 \hline
 Asia & india indian chinese china pakistan tamil sri muslim khan islamic\\ & ali state dynasty islam hindu government south temple asia muslims   \\
 \hline
 Education & school university college students education high state schools campus research \\ & science national student year center program institute medical public arts  \\
 \hline
 Politics & states government united state party law president national public u.s  \\ & court political rights act people election years international economic federal  \\
\end{tabular} 
\label{tab:mallet-topics}
\end{table*}

\begin{figure*}[t]
\centering
\includegraphics[width=\columnwidth]{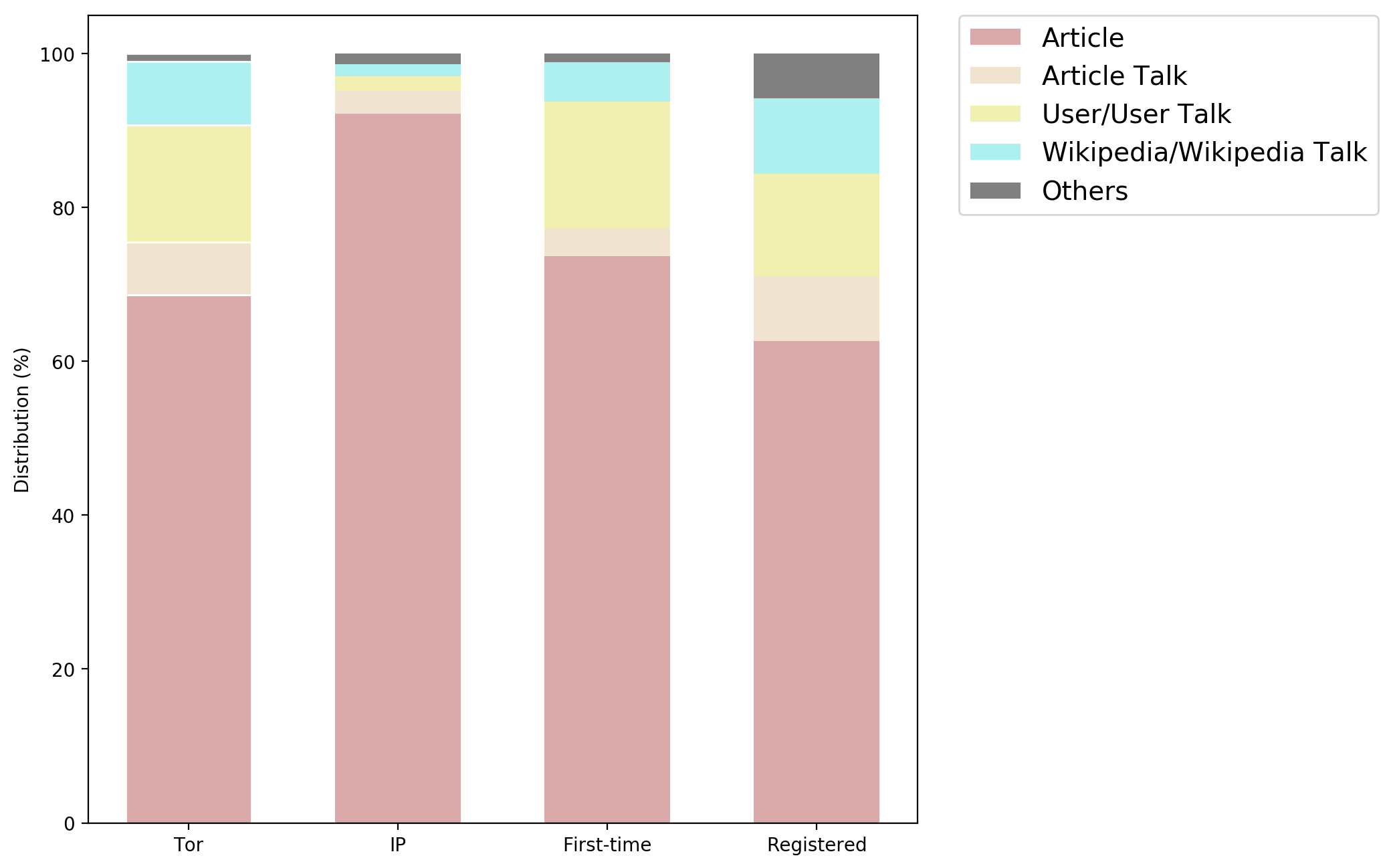}
\caption{Distribution of articles across namespaces for the four groups of edits.}
\label{fig:np_distribution}
\end{figure*}


\begin{figure*}[h!]
\centering\includegraphics[width=0.6\textwidth]{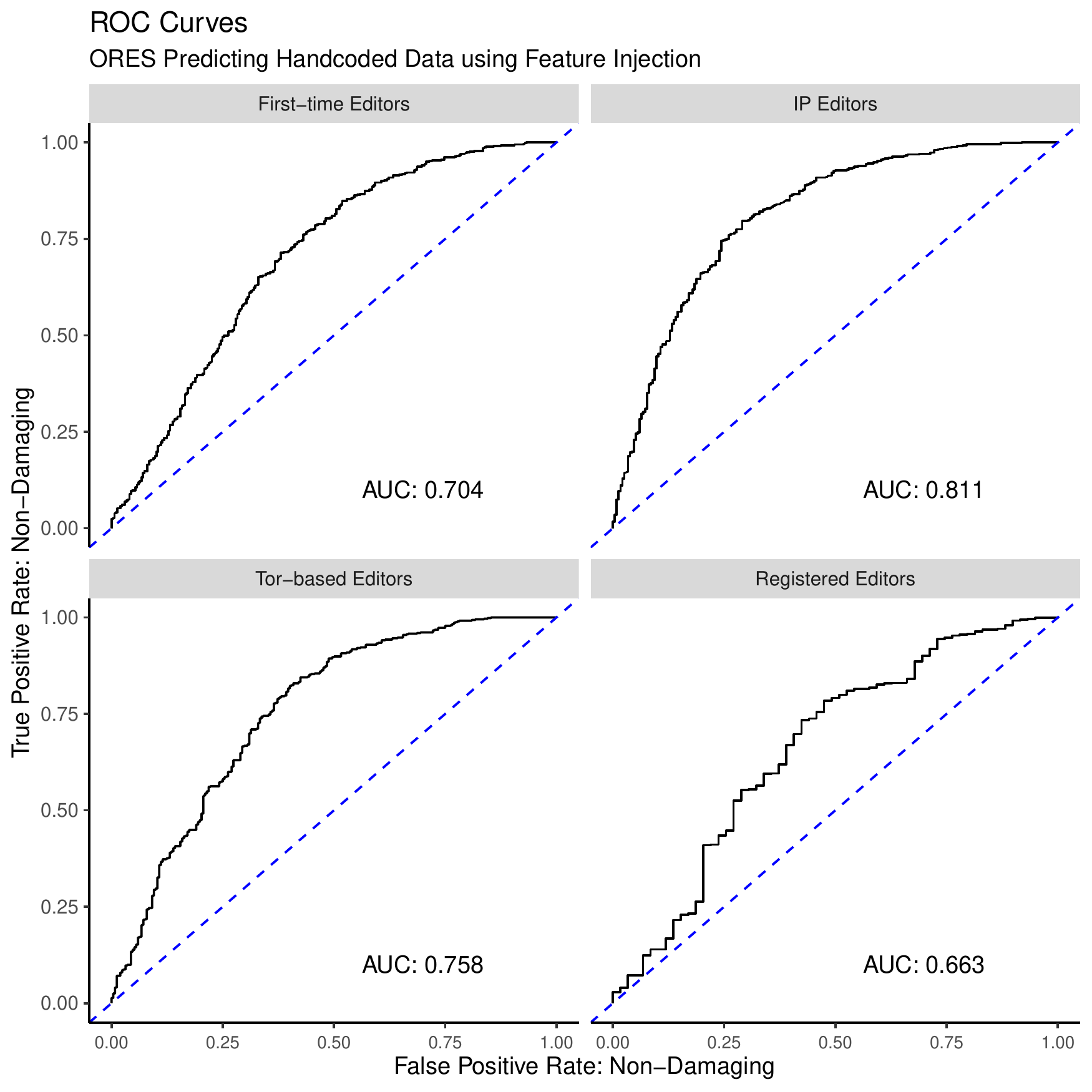}
\caption{ROC curve depicting false positive rate relative to the false negative rate for our sample user groups using the Non-Damaging model.
}
\label{fig:ROC_nd}
\end{figure*}

\begin{table*}
\caption{Tor users' activities on various Wikipedia language editions}
\centering
 \begin{tabular}{l c c} 
 \hline
 Language Editions & Total number of edits & Number of edits made by Tor users \\
 \hline
 German &  319,424,685 & 6,019  \\
 Russian & 67,743,927 & 3,795 \\
 Spanish & 78,601,767 & 2,343 \\
 French & 107,609,670 & 1,632 \\
 Chinese & 34,855,810 & 1,388 \\
 Polish & 48,483,852 & 456 \\
 Swedish & 41,298,921 & 437 \\
 Finnish & 16,377,486 & 261 \\
 Vietnamese & 36,846,744 & 179 \\
 Dutch & 95,223,918 & 141 \\
 
 \end{tabular}
 \label{tab:language-differences}
\end{table*}

\end{document}